%% file: main.tex
\documentclass[sigconf]{acmart}

\usepackage{booktabs}
\usepackage{graphicx}
\usepackage{xcolor}
\usepackage{colortbl}
\usepackage{multirow}
\usepackage{hyperref}
\usepackage{dblfloatfix} 
\usepackage{hhline}
\usepackage{subfig}
\usepackage{url}
\usepackage{mwe}

\definecolor{neworange}{RGB}{200,104,97}
\definecolor{newgreen}{RGB}{26,150,65}
\definecolor{newblue}{RGB}{44,123,182}
\definecolor{newpurple}{RGB}{94,60,153}
\definecolor{newbrown}{RGB}{166,97,26}

\acmDOI{10.475/123_4}

\acmISBN{123-4567-24-567/08/06}

\begin{document}
\title{Delineating Knowledge Domains in the Scientific Literature Using Visual Information}

\author{Sean T. Yang}
\affiliation{%
  \institution{University of Washington}
  \city{Seattle}
  \state{Washington}
}
\email{tyyang38@uw.edu}

\author{Po-shen Lee}
\affiliation{%
  \institution{University of Washington}
  \city{Seattle}
  \state{Washington}
}
\email{sephon@uw.edu}

\author{Jevin D. West}
\affiliation{%
  \institution{University of Washington}
  \city{Seattle}
  \state{Washington}
}
\email{jevinw@uw.edu}

\author{Bill Howe}
\affiliation{%
  \institution{University of Washington}
  \city{Seattle}
  \state{Washington}
}
\email{billhowe@cs.washington.edu}

\begin{abstract}
Figures are an important channel for scientific communication, used to express complex ideas, models and data in ways that words cannot. However, this visual information is mostly ignored in analyses of the scientific literature. In this paper, we demonstrate the utility of using scientific figures as markers of knowledge domains in science, which can be used for classification, recommender systems, and studies of scientific information exchange. We encode sets of images into a visual signature, then use distances between these signatures to understand how patterns of visual communication compare with patterns of jargon and citation structures. We find that figures can be as effective for differentiating communities of practice as text or citation patterns. We then consider where these metrics disagree to understand how different disciplines use visualization to express ideas. Finally, we further consider how specific figure types propagate through the literature, suggesting a new mechanism for understanding the flow of ideas apart from conventional channels of text and citations.  Our ultimate aim is to better leverage these information-dense objects to improve scientific communication across disciplinary boundaries.
\end{abstract}

%
%
%



\maketitle
\input{body}

\bibliographystyle{ACM-Reference-Format}
\bibliography{mybibliography}

\end{document}

%% file: body.tex
%

\section{Introduction}

Increased access to publication data has contributed to the emergence of the Science of Science (SciSci) as a field of study. SciSci studies metrics of knowledge production and the factors contributing to this production \cite{fortunato2018science}. Citations and text are the primary data types for measuring influence and tracking the evolution of scientific disciplines in this field. Dong et al. \cite{dong2017century} use citations to study the growth of science and observe the globalization of scientific development within the past century. Vilhena et al. \cite{Vilhena2014SocScience} characterize culture holes of scientific communication embedded in citation networks. However, among the studies in SciSci, the use of visualizationhas received little attention, despite being widely recognized as a significant communication channel within disciplines, across disciplines, and with the general public~\cite{lee2016viziometrics}. 

Humans perceive information presented visually better than textually\cite{nelson1976pictorial} due to the highly developed visual cortex\cite{ware2012information}. As a result, figures play a significant role in academic communication. The information density of a visualization or diagram can represent complex ideas in a compact form. For example, a neural network architecture diagram conveys an overview of the method used in a paper without requiring code listings or significant text. Moreover, the presence of a neural network diagram can be a better indicator that the paper involves the use of a neural network than any simple text features such as the presence of the phrase "neural network." 

Despite the importance of the figures in the scientific literature, they have received relatively little attention in the SciSci community. Viziometrics~\cite{lee2016viziometrics} is the analysis of visual information in the scientific literature. The term was adopted to distinguish this analysis from bibilometrics and scientometrics, while still conveying the common objectives of understanding and optimizing patterns of scientific influence and communication. Lee et al. \cite{lee2016viziometrics} has shown the relationship between visual information and the scientific impact of a paper. In this paper, we demonstrate that visual information can serve as an effective measure of similarity that can demarcate areas of knowledge in the scientific literature.

Different scientific communities use visual information differently and one can use these differences to understand communities of practice across traditional disciplines and show how ideas flow between these communities. 

We consider three hypotheses: H1) Sub-disciplines use distinguishable patterns of visual communication just as they use distinguishable jargon, H2) these patterns expose new modalities of communication that are not identifiable by either text or the structure of the citation graph, and H3) by classifying and analyzing use of specific types of figures, we can track the propagation and popularity of certain ideas and methods that are difficult to discern using text or citations alone (e.g., inclusion of neural network diagrams suggest contributions of new neural network architectures).

To test these hypotheses, we extract over 5 million scientific figures from papers on arXiv.org, process the images into low-dimensional vectors, then build a \emph{visual signature} for each field by clustering the vectors and computing the frequency distribution across clusters for each discipline. We use these signatures to reason about the similarity between fields, and compare these measures to prior work in understanding scientific community structure using text \cite{Vilhena2014SocScience} and the citation graph \cite{dreyfus1969appraisal,Vilhena2014SocScience}. Citations and text have been used to circumscribe knowledge domains, but this is the first study that shows that figures can also delineate fields. 
  
We compare the pairwise distances between these three matrices using the Mantel test \cite{mantel1967detection}, a common statistical test of the correlation between two distance matrices.  We find that the visual distance is moderately correlated to citation-based metrics (r = 0.706, p = 0.0001, z score = 5.103) and text-based metrics (r = 0.531, p=0.0002, z score = 5.019). We also perform hierarchical clustering on all distance matrices to provide a qualitative comparison of the results, finding that the hierarchical structure of the fields largely agrees, but with some significant exceptions.  We then consider pairs of fields that are visually distinct but similar in either text distance or citation distance, suggesting differences in the visual style of how ideas are presented.  For example, we find that \textit{Computation and Language} is visually distinct from other \textit{Computer Science} disciplines despite being quite similar in citation distance, because the former includes far more tables of data. 

Finally, we consider specific cases of the use of particular types of figures can indicates a common method or idea in a way that text and citation similarity do not.  We conduct a case study on two popular types of visualizations, neural network diagrams and embedding visualizations used to show clusters. The analysis indicates that visualizations can be used to make inferences about concept adoption within scientific communities. We also observe that the figures reveal the uptake of neural networks earlier than citation analysis, since citation counts take years to accrue. With this case study, we show the significance of visualizations in scientific literature, suggesting that the integration of figures into systems for bibilometric analysis, document summarization, information retrieval, and recommendation can improve performance and afford new applications.  Our focus is in the scientific literature, but our methods are directly applicable to other domains, including patents, web pages~\cite{akker2019vitor}, and news.

In this paper, we make the following contributions:
\begin{itemize}
    \item We present a method for delineating scholarly disciplines based on the figures and visualizations in the literature.
    \item We compare this method to prior results based on citations and text and find that different fields and sub-disciplines exhibit discernible patterns of visual communication (H1)
    \item  We find instances of fields that use similar jargon and cite similar sources, but are visually distinct, suggesting that visual patterns of communication are not redundant with other forms of communication (H2). 
    \item We present a method for identifying specific figure types and show that the presence of these figures in a paper can be used to understand concept adoption and a potential marker for tracking the evolution of scientific ideas (H3).
\end{itemize}

\section{Related Work}

Citations have been extensively studied and utilized as a measure of similarity among scientific publications. Marshakova proposed co-citation analysis \cite{marshakova1973co} which uses the frequency that papers are cited together as a measure of similarity. Citations are also utilized to delineate the emerging nanoscience fields in \cite{zitt2006delineating,leydesdorff2007nanotechnology} and are applied to design recommendation systems \cite{west2016IEEE}. However, citations only reveal the structural information with the scholarly literature and ignore the rich content in the articles.

Text has also received significant attention on analyzing the connection within scientific disciplines and documents, especially in citation recommendations \cite{huang2014refseer,strohman2007recommending}. Vilhena et al. \cite{Vilhena2014SocScience} proposed a text-based metric to characterize the jargon distance between disciplines. However, ambiguity and synonymity of text makes text-based model less ideal\cite{kuccuktuncc2012direction}.

Researchers have explored other aspects of a research paper for measuring the distance between disciplines. The frequency of mathematical symbols in papers are used to delineate fields by West et. al \cite{West2016JCDL}, but mathematical symbols are not as ubiquitous as other components. Visual communication is a significant channel for conveying scientific knowledge, but is relatively less explored.

A number of studies have focused on mining the scientific figures. Chart classification was well-studied by Futrelle et al. \cite{futrelle2003extraction}, Shao et al. \cite{shao2005recognition}, and Lee et al. \cite{lee2017phyloparser}. Recent studies have been focusing on the extraction of quantitative data from scientific visualizations, including line charts \cite{lu2007automatic,siegel2016figureseer}, bar charts \cite{al2015automatic}, and tables \cite{fang2012table}. Researchers have also investigated the techniques to understand the semantic messages of the scientific figures. Kembhavi et al. \cite{kembhavi2016diagram} utilized  a convolution neural network (CNN) to study the problem of diagram interpretation and reasoning. Elzer et al. \cite{elzer2011automated} studied the intended messages in bar charts. Several visualization-based search engines have also been presented. DiagramFlyer \cite{chen2015diagramflyer}, introduced by Chen et al., is a search engine for data-driven diagrams.  VizioMetrix\cite{lee2016viziometrix} and NOA\cite{charbonnier2018noa} are both scientific figures search engines with big scholar data, while they both work by examining the captions around the figures. We see visual-based models for demarcating knowledge domains as a next step in this area of research.  


\section{Method}

\subsection{Data}
The data for this study comes from the arXiv.  The arXiv is an open access repository for pre-prints in physics, mathematics, computer science, quantitative biology, quantitative finance, statistics, electrical engineering, systems science, and economics. 
The variety of disciplines allows consideration of information between fields, in contrast to more specialized repositories such as PubMed. There are 1,343,669 research papers which include 5,009,523 figures on arXiv through December 31st 2017. 

\subsection{Processing Pipeline}
\begin{figure}[!t]
    \centering
    \includegraphics[width=.8\linewidth]{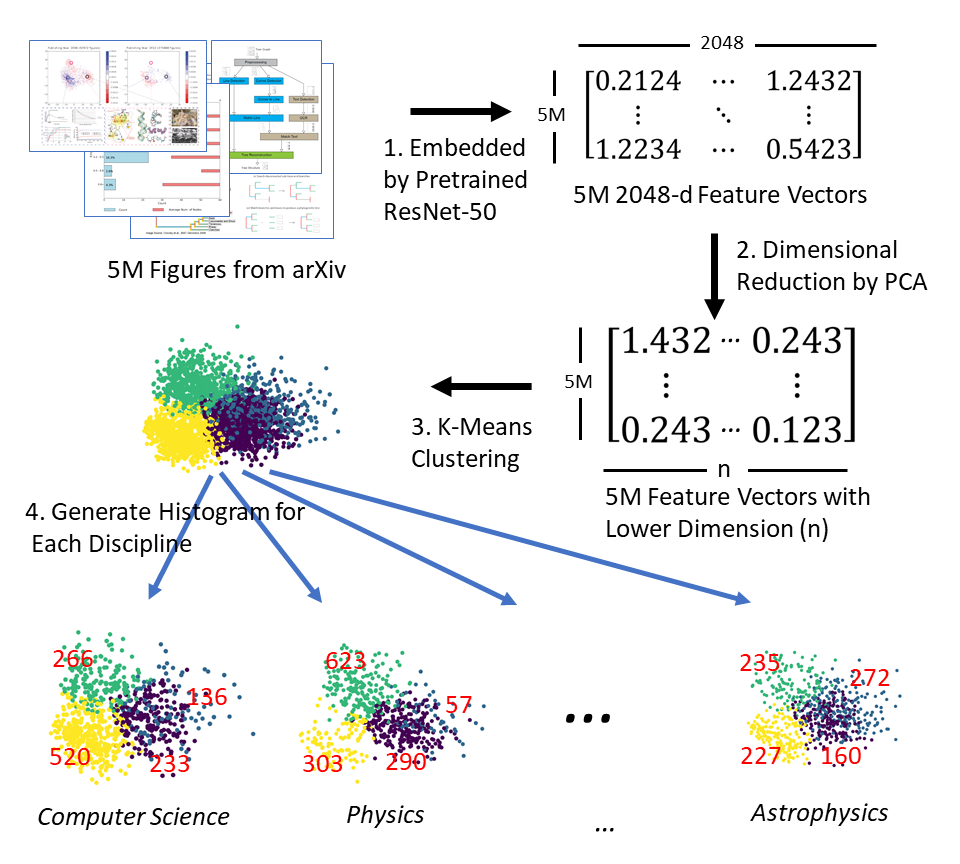}
    \caption{Overall pipeline.  Figures are mapped to vectors using ResNet-50, dimension-reduced, then organized into a histogram for each field.  The distances between these histograms are used to infer relationships and information flow.}
    \label{fig:pipeline}
\end{figure}

Fig. \ref{fig:pipeline} shows the pipeline to characterize scientific disciplines using visual information. Each step will be explained in the corresponding numbered paragraph.

\subsubsection{Convert Figures Into Feature Vectors}
We first embed each figure into a 2048-d feature vector using the pre-trained ResNet-50 \cite{he2016deep} model. The figures are re-sized and padded with white pixels to be 224 x 224 before being embedded by pre-trained ResNet-50. ResNet-50 was trained on the ImageNet \cite{deng2009imagenet} corpus of 1.2M natural images. Even though the model was trained on natural images, we find that the early layers of the network identify simple patterns (lines, edges, corners, curves) that are sufficiently general for the overall network to represent the combinations of edges and shapes that comprise artificial images as well. Although we posit that a custom neural network architecture could be designed to incrementally improve performance on artificial images, we do not further consider that direction in this paper.

\subsubsection{Dimension Reduction}
 We reduce the dimension of each figure vector using Principal Component Analysis (PCA). The high-dimensional vectors produced by ResNet-50 contain more information than is necessary for our application of computing the visual similarity between fields, and we seek to make the pipeline as efficient as possible. Plus, the ResNet model is pre-trained by natural images, while scientific figures have a lot more white areas, which make the embedding vectors more sparse, than natural images. Distances tend to be inflated in high dimensional space, reducing clustering performance~\cite{bellman2015adaptive}.  We follow the typical practice of applying dimension reduction prior to clustering.  Our original hypothesis was that a very low number of dimensions (10) would be sufficient to capture the differences between fields, but in our evaluation the higher values (200+) produced stronger correlations with other methods of delineating fields. We considered different values of this parameter using a sample of 1.5M figures from the 5M figure corpus. The results of the experiment are presented in Section \ref{result0}. 


\subsubsection{Cluster the Figure Corpus} \label{kmean}
The distribution of different types of figures carries significant information about how the visual communication is different in each discipline and could further represent each category. We cluster our figure corpus with K-Means clustering to aggregate similar figures. Although more advanced methods of clustering could provide better results, we aim to demonstrate that the approach can work even with very simple methods.  The objective of this paper is to show the utility of the figures for potential applications, rather than to propose a specialized framework for specific task. The experimental results are shown in Section \ref{result0}.


\subsubsection{Visual Signatures for Each Discipline}
We cluster the figures with number of centroid \(k = 4\) and generate the normalized histogram for each discipline to acquire visual signature of each discipline.


After the visual signature of each discipline is generated, we calculate the euclidean distance between each pair of disciplines. We evaluate the computed visual similarity between disciplines by comparing to citation-based and text-based metrics described in previous work, which are explained in Section \ref{compare}.

\subsection{Classifying Figure Types}

\begin{figure}[ht]
   \subfloat[\label{fig:NN_diagram_example}]{%
      \includegraphics[ width=0.70\columnwidth]{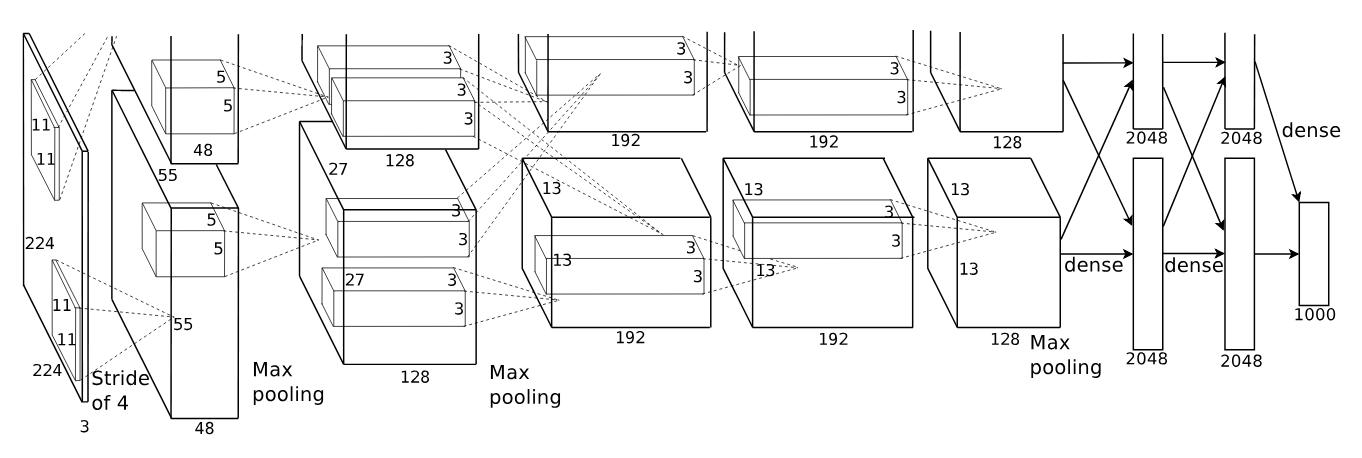}}
\hspace{\fill}
   \subfloat[\label{fig:CEV_example} ]{%
      \includegraphics[ width=0.25\columnwidth]{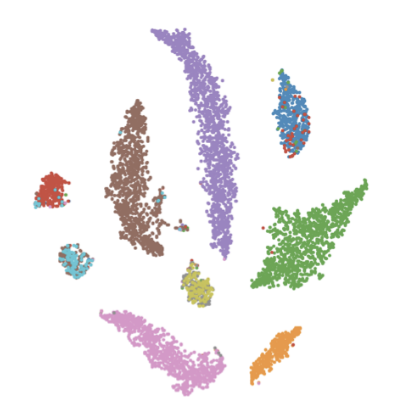}}
\caption{\label{fig:examples} Examples of neural network diagram and embedding visualization. (a) An example of neural network diagram. The diagram is borrowed from AlexNet paper~\cite{krizhevsky2012imagenet}. (b) An example of embedding visualization. The plot is borrowed from MultiDEC paper \cite{yang2019multidec}.}
\end{figure}

\begin{figure}[]
    \centering
    \includegraphics[width=0.8\linewidth]{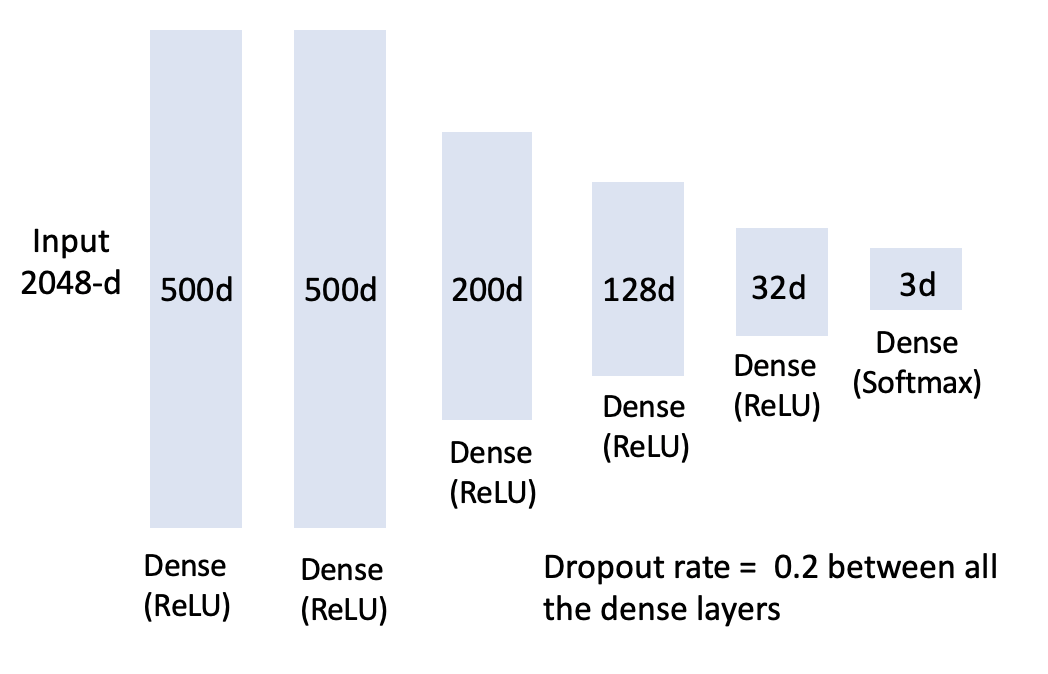}
    \caption{The architecture of the neural network diagrams and embedding visualization classifier.}
    \label{fig:classifier_architecture}
\end{figure}

In this section, we describe the process to train the classifier to identify specific figure types, which we will use to understand how the use of particular styles of visualization and diagrams propagate through the literature. We consider two specific examples: neural network diagrams (associated with the rapid increase of neural network methods in the literature) and clustering plots (associated with the use of unsupervised learning). Examples of these visualizations are shown in Figure \ref{fig:examples}. 
Sethi et al. \cite{sethi2018dlpaper2code} characterize six different figure types to demonstrate neural network architecture. We label 10,651 figures from arXiv, which includes 1,503 neural network diagrams, 1,057 embedding visualizations, 8,091 negative examples. For neural network diagrams, we label them according to the taxonomy suggested by Sethi et al. \cite{sethi2018dlpaper2code}, but we exclude figures in table format. We consider a figure as an embedding visualization if the figure is used to visualize the representation distribution of the data. The annotators make use of images and captions to label the images. We extract visual features from the fully connected layer of a ResNet-50\cite{he2016deep} model, which is pre-trained by 1M ImageNet dataset\cite{deng2009imagenet}. The figures are resized to 224x224 and a 2048-d numeric vector is acquired for each figure. The labeled image set is then split into training, validation, and test set with 8:1:1 ratio to train a deep neural network (DNN) classifier. We tune the depth of the model, dimension of the layers, dropout rate, learning rate, decay ratio, and training epochs. The architecture of the final model is shown in Figure \ref{fig:classifier_architecture} and implementation details is shown in Table \ref{table:implementation}. 

\begin{table}[]
\caption{Implementation details for training the neural network diagrams and embedding visualization classifier.}
\label{table:implementation}
\begin{tabular}{c|c|c|c|c}
\begin{tabular}[c]{@{}c@{}}Learning\\ Rate\end{tabular} & Decay & Epoch & Batch Size & Loss                                                                 \\ \hline
0.001                                                   & 0.001 & 150   & 256        & \begin{tabular}[c]{@{}c@{}}Categorical \\ Cross Entropy\end{tabular}
\end{tabular}
\end{table}

\section{Comparison with Citation- and Text-based Methods} \label{compare}

We use the Mantel test \cite{mantel1967detection}, a standard statistical test of the correlation between two matrices, to compare visual distance with the distance matrices created by (1) Average shortest citation distance \cite{dreyfus1969appraisal,Vilhena2014SocScience} and (2) Natural language jargon distance~\cite{Vilhena2014SocScience}. Citations and text have been extensively analyzed and employed to measure the similarity among research articles, and both of the measures have had success on information retrieval and recommendation systems among scholarly documents. Therefore, we consider citation distance as our benchmark of the task and text distance as alternative comparison. 


\subsection{Average Shortest Citation Path}


We compute the average shortest path between each pair of fields as a measure of similarity. Average shortest path \cite{dreyfus1969appraisal} is one of the three most robust measures \cite{boccaletti2006complex} of network topology, in addition to its clustering coefficient and its degree distribution. Vilhena et al \cite{Vilhena2014SocScience} used this method to measure distance in the citation network to compare with their text-based metric. 

Average shortest path is computed as follows:
\[
D_{ij} = \frac{1}{n_in_j}\sum_{n_i} \sum_{n_j} d(v_i,v_j)
\]
where \(n_i\) is the number of vertices in field \(i\) and \(n_j\) is the number of vertices in field \(j\). The average shortest path between field \(i\) and field \(j\), \(D_{ij}\), is the average of all paths between all vertex pairs, \(v_i\) and \(v_j\). 

Our citation graph is obtained from the SAO/NASA Astrophysics Data System (ADS)\cite{eichhorn1994overview}, a digital library portal maintaining three bibliographic databases containing more than 13.6 million records covering publications in Astronomy and Astrophysics, Physics, and the arXiv e-prints. The creation of the citations in ADS \cite{accomazzi2006creation} is started by scanning the full-text of the paper to retrieve bibcode for each reference string in the article, followed by computing the similarity score between the ADS record and the bibcode. The citation pairs are generated if the similarity is higher than the threshold. This data has been extensively used on several bibliographic studies \cite{garfield2006history,kurtz2017measuring}. There are 14,555,820 citation edges within our arXiv data corpus. 

\subsection{Jargon Distance}

We also compare our results to text metrics based on cultural information as represented by patterns of discipline-specific jargon. Jargon distance was first proposed by Vilhena et al. \cite{Vilhena2014SocScience}, where the authors quantitatively measure the communication barrier between fields using n-grams from full text. The jargon distance (\(E_{ij}\)) between field \(i\) and field \(j\) is defined as the ratio of (1) the entropy \(H\) of a random variable \(X_i\) with a probability distribution of the jargon or mathematical symbols within field \(i\) and (2) the cross entropy \(Q\) between the probability distributions in field  \(i\) and field \(j\): 
\[
E_{ij} = \frac{H(X_i)}{Q(p_i||p_j)} = \frac{-\sum_{x\in X}p_i(x)\log_2p_i(x)}{-\sum_{x\in X}p_i(x)\log_2p_j(x)}
\]

Imagine a writer from field \(i\) trying to communicate with a reader from field \(j\). The writer has a codebook \(P_i\) that maps the natural language or mathematical symbols to codewords that the reader has to decode using the codebook \(P_j\) from field \(j\). A small jargon distance means high communication efficiency between two fields and are closely related. This metric could be easily applied to natural language jargon to explore how the communication varies through these two channels across disciplines. We compute the jargon distance between two different disciplines by applying the metrics on unigram from abstracts. 

\section{Results}

We show that the distance between visual signatures can be used to determine the overall relationships between fields in a manner similar to prior methods, but that this approach also exposes information that prior methods cannot. In Section \ref{result0}, we present the experimental results on picking the number of dimensions and clusters. In Section \ref{result1}, we show the capacity of visual distance to reveal the relationships across scientific disciplines by showing global agreement between visual distance and citation distance (H1). In Section \ref{result2}, we examine each cluster to understand the visual composition and find that each cluster is dominated by a certain type of visualization, extending prior work in the life sciences that used coarse-grained labeling of figure types~\cite{lee2016viziometrics}.
In Section \ref{result3}, we show that citation distance and visual distance disagree in certain cases, and consider one case in particular (H2).  
Finally, we consider cases where the presence of a particular type of figure can indicate the use of a method or concept in a way that text and citation similarity do not in Section \ref{result4} (H3). We demonstrate that the figures in the scientific literature can serve as an indicator of concept adoption that travels faster than citation count.

\subsection{Choosing the number of dimensions and clusters} \label{result0}
\begin{table*}[hbt]
\centering
\caption{Choosing the number of clusters (k).}
\label{tab:PCA}
\resizebox{0.7\textwidth}{!}{%
\begin{tabular}{c|c|c|c|c}
Dimension & \begin{tabular}[c]{@{}c@{}}Explained Variance\\ Ratio\end{tabular} & \begin{tabular}[c]{@{}c@{}}Average of Correlations\\ to Citation Distance\end{tabular} & \begin{tabular}[c]{@{}c@{}}Maximum Correlation\\  to Citation Distance\end{tabular} & \begin{tabular}[c]{@{}c@{}}Maximum\\  at k=?\end{tabular} \\ \hline
16        & 52.0\%                                                             & 0.661                                                                                  & 0.737                                                                               & 15                                                        \\ \hline
32        & 63.7\%                                                             & 0.631                                                                                  & 0.768                                                                               & 3                                                         \\ \hline
64        & 73.9\%                                                             & 0.660                                                                                  & 0.769                                                                               & 4                                                         \\ \hline
128       & 82.3\%                                                             & 0.662                                                                                  & 0.770                                                                               & 4                                                         \\ \hline
256       & 88.9\%                                                             & 0.672                                                                                  & 0.793                                                                               & 4                                                         \\ \hline
320       & 90.7\%                                                             & 0.674                                                                                  & 0.793                                                                               & 4                                                        
\end{tabular}
}
\end{table*}

Our pipeline involves two hyperparameters: the number of dimensions to retain via PCA and the number of clusters to assume when constructing visual signatures. We determine these parameters experimentally.  The results of our analysis of PCA dimensions appear in Table \ref{tab:PCA}. The explained variance ratio shows the percentage of variance explained by the selected components. The variance explained grows insignificantly after 256 components. The average correlation with citation distance shows the average of the  correlations between visual distance and citation distance across all the numbers of centroid \(k\) (from 2 to 30). We evaluate our method by conducting the Mantel test \cite{mantel1967detection} to compare the correlation between visual distance and citation distance. It confirms our hypothesis that the correlation increases when more components are used, but it converges after sufficient information is preserved. Maximum correlation to citation distance shows the maximum correlation of the specified dimension among different options of number of centroid \(k\), and the \(k\) contributing the maximum correlation is shown in ''Maximum at k = ?''. Surprisingly, the maximum correlation happens at larger number of centroid with low dimension of figure vector. Our interpretation is that there is not sufficient information preserved by low dimensional space.

We ran a second experiment to determine the number of centroids \(k\). Initially, we expected the correlation with other measures to be higher using larger values of $k$, since the diversity of figures in the literature appears vast.  However, considering k = 100, 200, and 400, we found that larger values  of \(k\) generate lower correlations with citation distance (correlation coefficient around 0.4), due to overfitting to rare, low-confidence clusters.  Lowering $k$ to the range of 2 to 30 performed better; these results appear in Table \ref{tab:PCA}.  The relatively low values of $k$ suggest that there are relatively few modalities of visual communication in use across fields.  The maximum correlation occurred at k = 4 in most of the experiments. We further discuss the interpretation of these results in Section \ref{result2}.

\subsection{Delineating Disciplines} \label{result1}

In this section, we demonstrate the ability of visual distance to characterize the relationships between fields, quantitatively and qualitatively. Quantitatively, we conduct the Mantel test \cite{mantel1967detection} with Spearman rank correlation method to compare two different distance matrices to reveal the similarity between two structures. We also perform hierarchical clustering using UPGMA algorithm \cite{rohlf1968tests} to visualize the hierarchical relationships across disciplines, qualitatively. Vilhena et al.\cite{Vilhena2014SocScience} used similar technique to qualitatively visualize how disciplines are delineated, but the data they used was from JSTOR, which focuses on biological science and social science so that it is not comparable with our task.

\begin{table}[!b]
\centering
\caption{The correlation results between distance matrices.}
\label{tab:corr}
\begin{tabular}{c|c|c}
\multicolumn{1}{l|}{} & \multicolumn{1}{l|}{} & Results                                                                    \\ \hline
Visual Distance       & Citation Distance     & \begin{tabular}[c]{@{}c@{}}r = 0.706\\ p = 0.0001\\ z = 5.103\end{tabular} \\ \hline
Visual Distance       & Jargon Distance       & \begin{tabular}[c]{@{}c@{}}r = 0.531\\ p = 0.0002\\ z = 5.019\end{tabular} \\ \hline
Jargon Distance       & Citation Distance     & \begin{tabular}[c]{@{}c@{}}r = 0.697\\ p = 0.0001\\ z = 5.989\end{tabular}
\end{tabular}
\end{table}

Table \ref{tab:corr} shows the correlation results between different distances. The first two columns indicate the methods being compared and the Results column shows the correlations. The correlation between visual distance and citation distance (\(r = 0.706\), p value = 0.0001, z score = 5.103) is higher than the correlation between jargon distance and citation distance (\(r = 0.697\), p value = 0.0001, z score = 5.989), providing evidence for our hypothesis that styles of visual communication are a stronger indicator of communication and influence than the terminology used by a field. Visual distance is also moderately correlated to jargon distance with \(r = 0.531 \), p value = 0.0002, and z score = 5.019. This result is expected. It verifies our first hypothesis: sub-disciplines use distinguishable patterns of visual communication. Correlation between visual distance and citation distance is sufficient enough to show that visual distance is capable of characterizing general relationships between disciplines, but it also reveals that there are still differences between citation distances and visual distance. We will elaborate the different connections visual distance expose in Section \ref{result2}.


\begin{figure*}[!ht]
    \centering
    \includegraphics[width=.9\linewidth]{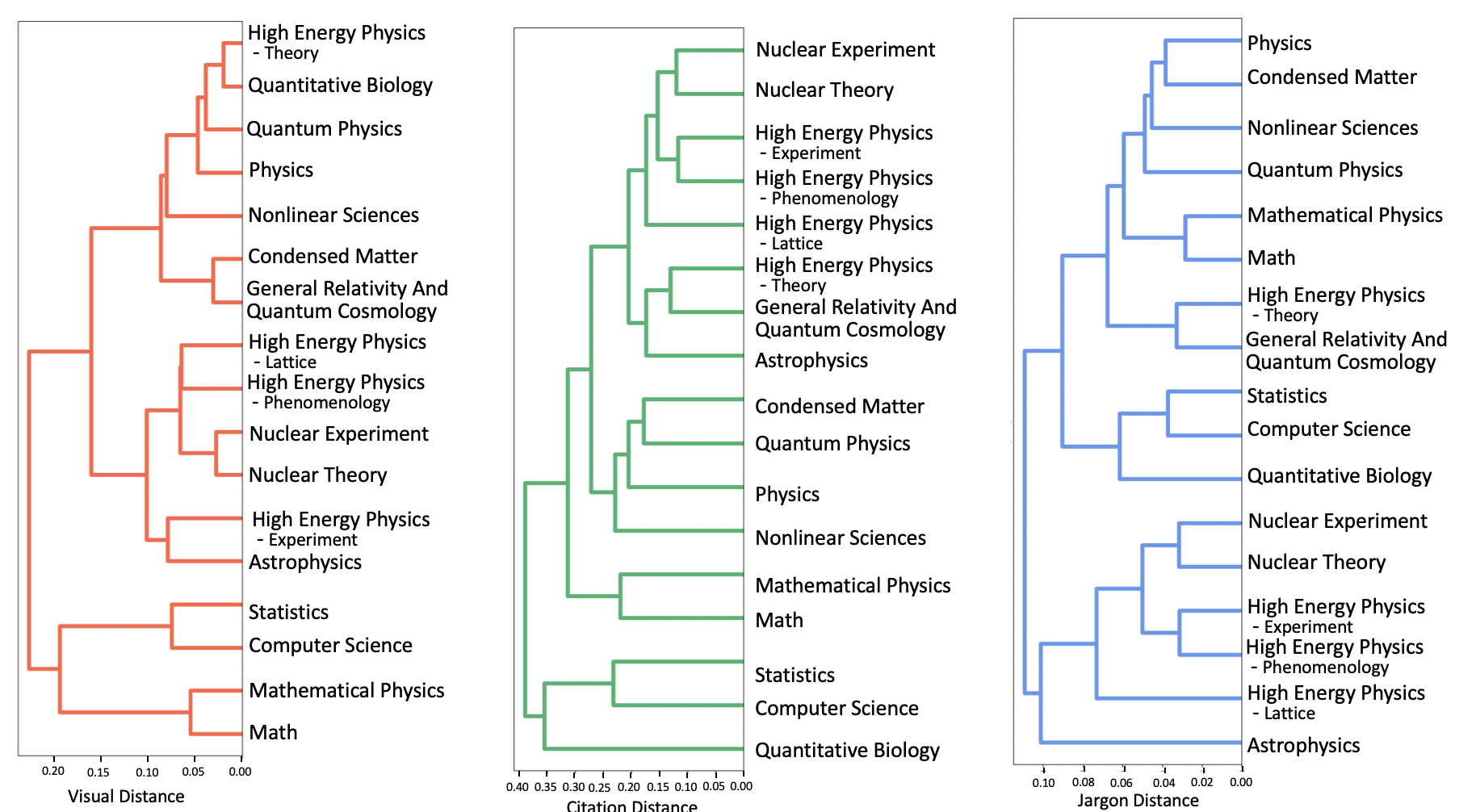}
    \caption{The hierarchical clustering dendrogram of visual distance (left), citation distance (middle), and jargon distance (right). Citation distance is a benchmark in our task. It shows similar pattern as visual distance where \textit{Computer Science}, \textit{Statistics}, \textit{Math}, and \textit{Mathematical Physics} are separated from the rest of the disciplines. The inconsistency between citation distance and visual distance is \textit{Quantitative Biology}, which is clustered with physics-related disciplines in visual distance while it is isolated in citation distance. On the other hand, Jargon distance segregates disciplines differently from visual distance and citation distance in the high level. High Energy Physics and Nuclear are separated from the rest where Quantitative Biology, Computer Science and Statistics are isolated in the sub-cluster.}
    \label{fig:dendrogram_triple}
\end{figure*}
We then perform hierarchical clustering, using the UPGMA algorithm \cite{rohlf1968tests}, to qualitatively visualize how different methods group similar disciplines together and separate dissimilar disciplines. The hierarchical clustering results for visual distance, citation distance, and jargon distance are shown in Fig.\ref{fig:dendrogram_triple}. We observe similar patterns between visual distance and citation distance where \textit{Computer Science}, \textit{Statistics}, \textit{Math}, and \textit{Mathematical Physics} are isolated from other physics-related fields of study. There is inconsistency between visual distance and citation distance in the field of  \textit{Quantitative Biology}, which is the outlier in citation distance, but is assigned to the physics-related cluster in visual distance.

\subsection{Analyzing Clusters}  \label{result2}

\begin{figure}[t]

    \centering
    \includegraphics[width=.8\linewidth]{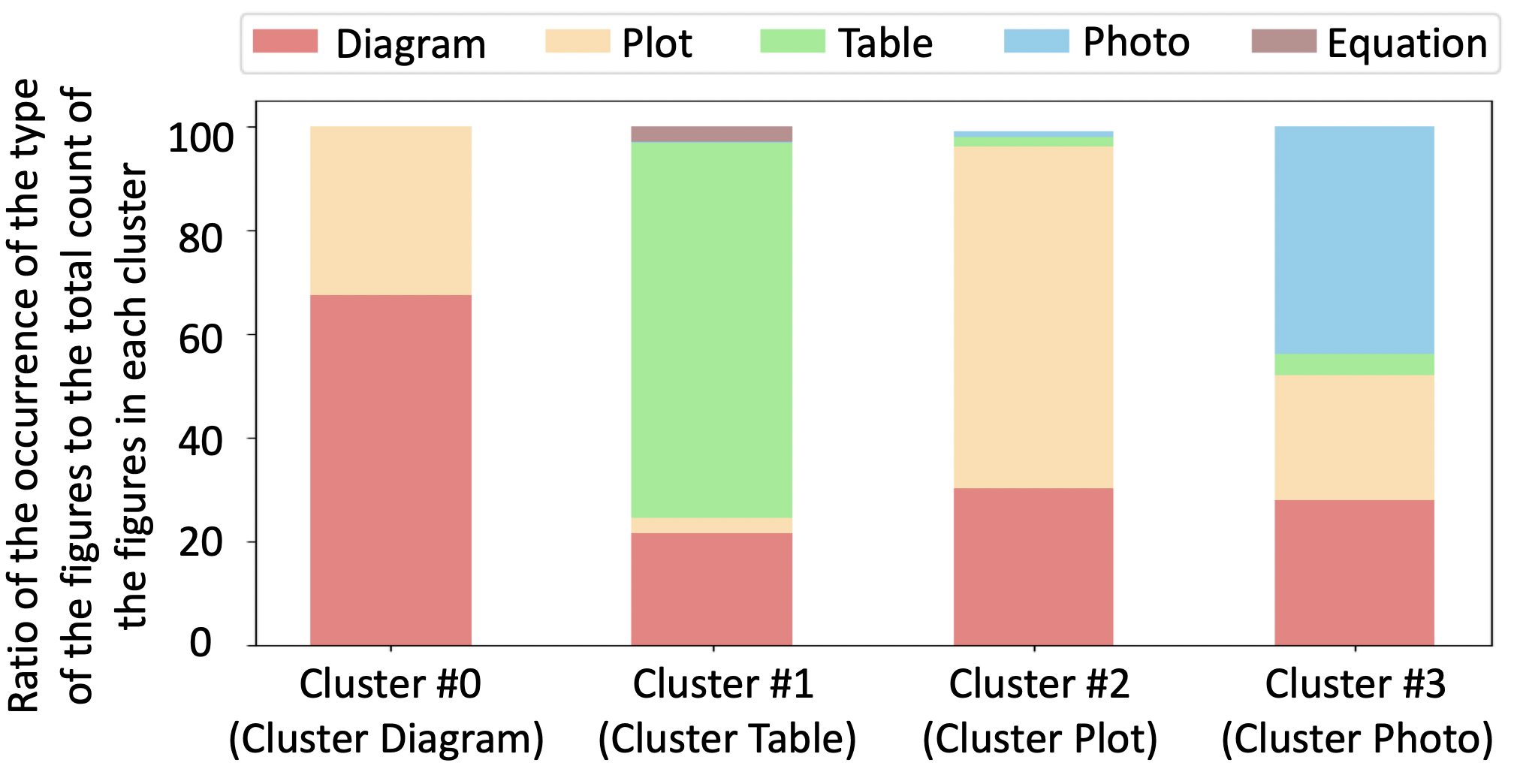}
    \caption{The visual composition of each cluster. It appears that each cluster has one dominant visualization.}
    \label{fig:cluster_profile}

\end{figure}

We classify the figures in each cluster to understand the visual composition of each cluster. We use the convolutional neural network classifier in \cite{lee2017phyloparser} to categorize figures into five categories: (1) Diagrams (2) Plots (3) Table (4) Photo and (5) Equation. The classification results are shown in Fig. \ref{fig:cluster_profile}. Surprisingly, each cluster is prominently associated with a certain type of visualization: Cluster\#0 is primarily composed of diagrams (\textsf{Diagram}), Cluster\#1 is primarily composed of tables \textsf{Table}, Cluster\#2 is primarily composed of plots of quantitative information (\textsf{Plot}, and Cluster\#3 is primarily composed of photos \textsf{Photo}. These results corroborate previous work that used supervised methods and manual labeling to categorize figures into five classes (Diagram, Plot, Table, Photo, and Equation)~\cite{lee2016viziometrics}. 
The distribution of figures helps to reveal the properties of each discipline. For instance, Cluster \textsf{Plot} is dominant in \textit{Quantitative Biology} (48\%) and \textit{Nuclear Experiment} (60\%), which may indicate the degree to which these fields can be considered experimental and data-driven. The distribution could further be used to group similar disciplines and separate the dissimilar fields as we show in the previous section.

\subsection{Visuals delineate differently than citations}  \label{result3}
In this section, we focus on the cases in computer science where visual distance and citation distance disagree and we validate our second hypothesis: visual patterns expose new modalities of communication that are not identifiable by either text or the structure of the citation graph. The analysis aims to answer the following questions: (1) Where are there visual differences in the disciplinary landscape when compared to citation differences? (2) What is revealed about the fields where visual differences occur?

\begin{figure}[t!]

    \centering
    \includegraphics[width=\linewidth]{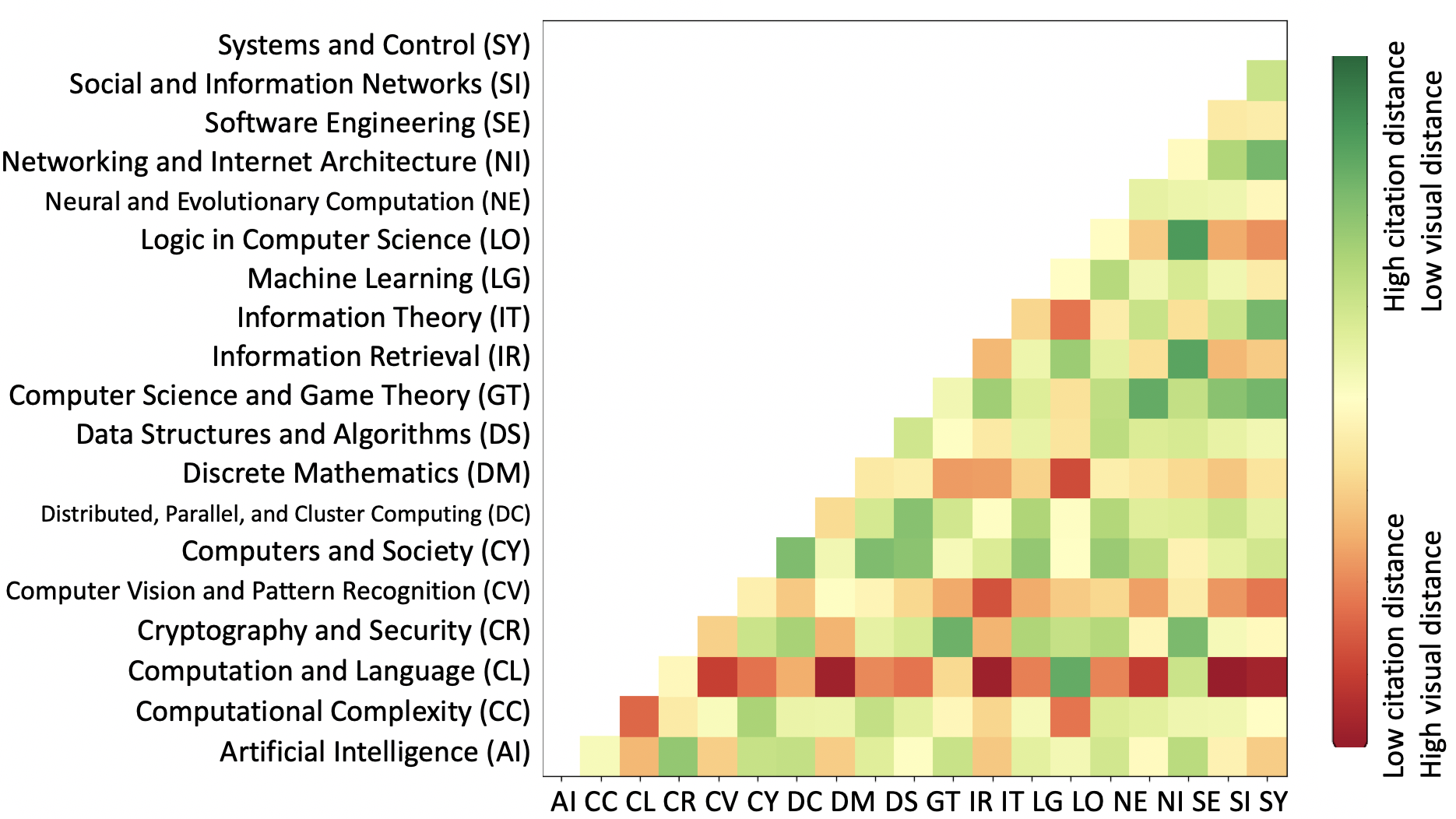}
    \caption{Heat map of differences between visual and citation distance. We normalize visual distance and citation distance and subtract visual distance from citation distance to expose the discrepancies. Red indicates that two subfields are visually distant but near in citation distance. Green indicates that two subfields are distinct in citation distance but visually similar.  \textit{Computation and Language} is visually different across the subfields in \textit{Computer Science} but relatively close in terms of citation distance. }
    \label{fig:heatmap_vs_cd_cs}

\end{figure}

We normalize visual distance and citation distance, then subtract visual distance from citation distance to expose the discrepancies. Fig. \ref{fig:heatmap_vs_cd_cs} shows that there is a significant disagreement between visual distance and citation distance for the subfield \textit{Computation and Language}.  Red cells show the disagreements where fields are visually distinct but similar in citation distance. Green cells, in contrast, indicate disciplines that are visually similar, but far apart in citation distance. We observe that \textit{Computation and Language} is generally close to all other categories in \textit{Computer Science}, but visually distinct.  We further examine the visual profile of \textit{ Computation and Language} in order to better understand the reasons for the divergence between these two distances.

\begin{figure}[hbt]

    \centering
    \includegraphics[width=.9\linewidth]{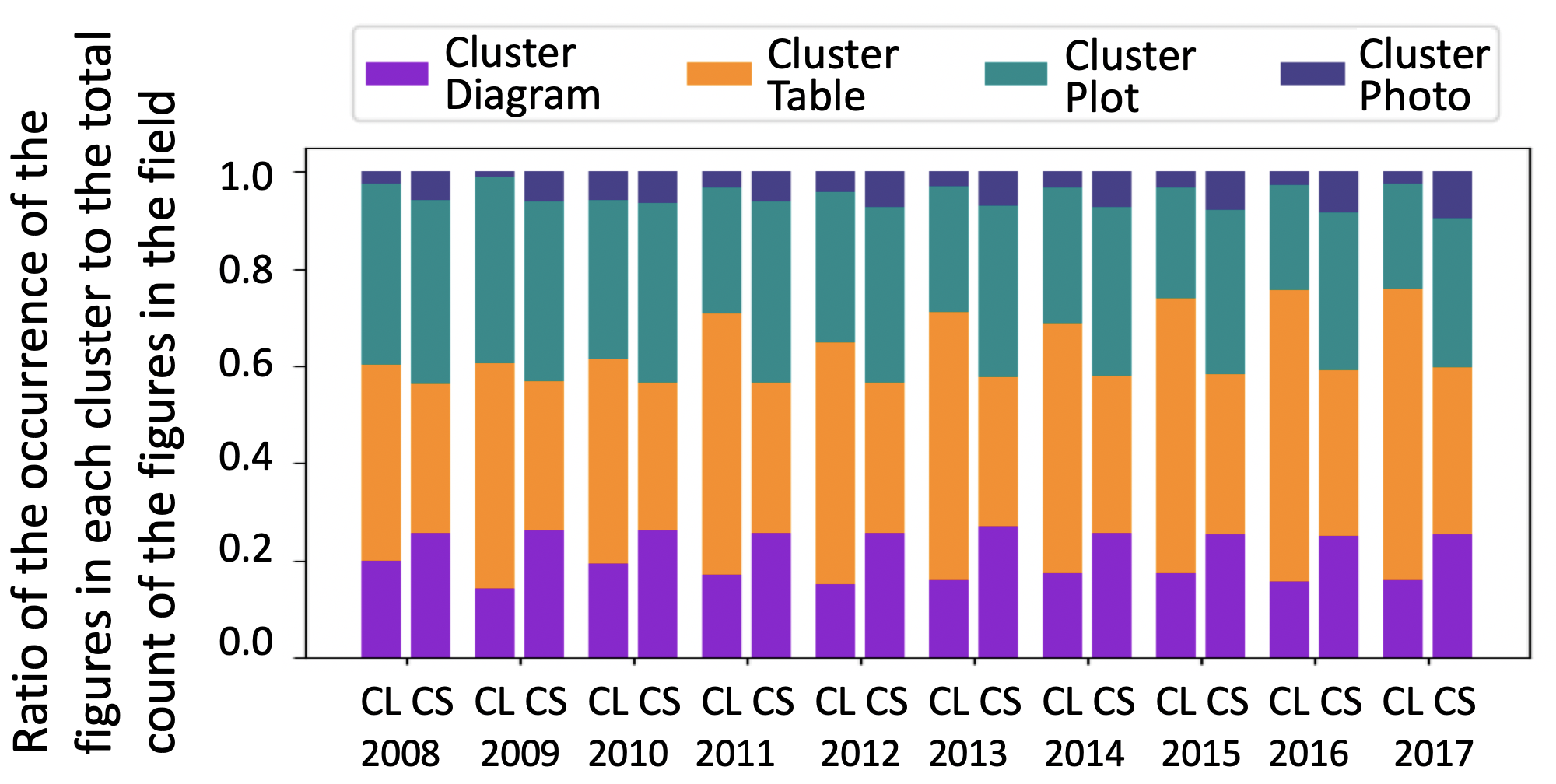}
    \caption{The chart shows how the distribution of the clusters evolves in \textit{Computation and Language} and \textit{Computer Science} over the past ten years. We could observe that Cluster Table has been growing in \textit{Computation and Language} and researchers in \textit{Computation and Language} use a relatively low number of figures in the Photo Cluster.}
    \label{fig:stacked_cl}

\end{figure}

Fig. \ref{fig:stacked_cl} shows the distribution of the figure usage in \textit{Computation and Language} (CL) and \textit{Computer Science} (CS) over the past ten years. We make two observations from this stacked bar chart: (1) Cluster \textsf{Table} dominates the visual communication style with over 50\% in \textit{Computation and Language} in 2017, compared to approximately 30\% in \textit{Computer Science}, and it has been growing over the past few years.  (2) The researchers in \textit{Computation and Language} use very few figures associated with Cluster \textsf{Photo}. 
We further investigate the reason that tables are largely used in \textit{Computation and Language} by analyzing the cluster textually. We conduct topic modeling on the captions of the figures of Cluster Table using Non-negative Matrix Factorization (NMF) \cite{lee1999learning} with five topic numbers. In Table \ref{tab:topic_cl}, we display the top 10 keywords of each topic along with the ratio of the count of the figures in each topic to the total count in the cluster over the past 10 years. We also look at the images in each topic to help us understand the purpose of each topic. Based on the keywords and the images, we can infer that Topic 0 mostly contains table with comparison data to other models, Topic 1 includes the examples of the language and words, Topic 2, which is similar to Topic 0, also involves comparing results between different models. Topic 3 consists of statistics about the dataset. Topic 4 is a mix of the tables and diagrams which mostly are used to illustrate the architecture of LSTM models. It appears that tables to compare the accuracy of different models have been growing significantly, from 46.4\% (28.6\% + 17.8\%) in 2008 to 60\% (47.6\% + 12.4\%) in 2017, suggesting that an empirical regime of research is dominant, perhaps due to improved access to advanced computational infrastructure, easy access to data and code, and the rapid growth of the field itself.

\begin{table}[!hbt]
\centering
\caption{Top 10 keywords for each topic in Cluster Table along with the ratio of the figure in each topic over time.}
\label{tab:topic_cl}
\resizebox{\columnwidth}{!}{%
\begin{tabular}{cccccc}
\multicolumn{6}{c}{}                                                                                                                                                                                                                                                                                                                                                                                                                                                                                                                                                                                                                                                                                                                                                                                                                                 \\
\multicolumn{1}{c|}{}                                                                 & \multicolumn{1}{c|}{\textbf{Topic 0}}                                                                                                                             & \multicolumn{1}{c|}{\textbf{Topic 1}}                                                                                                                 & \multicolumn{1}{c|}{\textbf{Topic 2}}                                                                                            & \multicolumn{1}{c|}{\textbf{Topic 3}}                                                                                                            & \textbf{Topic 4}                                                                                                                           \\ \hline
\multicolumn{1}{c|}{\textbf{\begin{tabular}[c]{@{}c@{}}Cluster\\ Table\end{tabular}}} & \multicolumn{1}{c|}{\begin{tabular}[c]{@{}c@{}}results\\ table\\ models\\ different\\ performance\\ best\\ scores\\ dataset\\ comparison\\ accuracy\end{tabular}} & \multicolumn{1}{c|}{\begin{tabular}[c]{@{}c@{}}words\\ figure\\ word\\ number\\ example\\ table\\ sentence\\ example\\ sentences\\ used\end{tabular}} & \multicolumn{1}{c|}{\begin{tabular}[c]{@{}c@{}}et\\ al\\ 2015\\ 2016\\ 2014\\ 2017\\ 2013\\ results\\ 2011\\ taken\end{tabular}} & \multicolumn{1}{c|}{\begin{tabular}[c]{@{}c@{}}set\\ test\\ training\\ data\\ development\\ sets\\ table\\ dev\\ used\\ statistics\end{tabular}} & \begin{tabular}[c]{@{}c@{}}model\\ language\\ trained\\ baseline\\ lstm\\ proposed\\ models\\ attention\\ layer\\ performance\end{tabular} \\ \hline
\rowcolor[HTML]{656565} 
\multicolumn{1}{|c|}{\cellcolor[HTML]{656565}{\color[HTML]{FFFFFF} year}}             & \multicolumn{1}{c|}{\cellcolor[HTML]{656565}{\color[HTML]{FFFFFF} ratio}}                                                                                         & \multicolumn{1}{c|}{\cellcolor[HTML]{656565}{\color[HTML]{FFFFFF} ratio}}                                                                             & \multicolumn{1}{c|}{\cellcolor[HTML]{656565}{\color[HTML]{FFFFFF} ratio}}                                                        & \multicolumn{1}{c|}{\cellcolor[HTML]{656565}{\color[HTML]{FFFFFF} ratio}}                                                                        & \multicolumn{1}{c|}{\cellcolor[HTML]{656565}{\color[HTML]{FFFFFF} ratio}}                                                                  \\ \hline
\multicolumn{1}{|c|}{2008}                                                            & \multicolumn{1}{c|}{28.6\%}                                                                                                                                       & \multicolumn{1}{c|}{25.0\%}                                                                                                                           & \multicolumn{1}{c|}{17.8\%}                                                                                                      & \multicolumn{1}{c|}{22.9\%}                                                                                                                      & \multicolumn{1}{c|}{5.7\%}                                                                                                                 \\ \hline
\multicolumn{1}{|c|}{2009}                                                            & \multicolumn{1}{c|}{31.1\%}                                                                                                                                       & \multicolumn{1}{c|}{26.8\%}                                                                                                                           & \multicolumn{1}{c|}{16.1\%}                                                                                                      & \multicolumn{1}{c|}{18.6\%}                                                                                                                      & \multicolumn{1}{c|}{7.4\%}                                                                                                                 \\ \hline
\multicolumn{1}{|c|}{2010}                                                            & \multicolumn{1}{c|}{31.2\%}                                                                                                                                       & \multicolumn{1}{c|}{24.2\%}                                                                                                                           & \multicolumn{1}{c|}{16.9\%}                                                                                                      & \multicolumn{1}{c|}{21.0\%}                                                                                                                      & \multicolumn{1}{c|}{6.7\%}                                                                                                                 \\ \hline
\multicolumn{1}{|c|}{2011}                                                            & \multicolumn{1}{c|}{34.2\%}                                                                                                                                       & \multicolumn{1}{c|}{25.1\%}                                                                                                                           & \multicolumn{1}{c|}{17.2\%}                                                                                                      & \multicolumn{1}{c|}{16.3\%}                                                                                                                      & \multicolumn{1}{c|}{7.2\%}                                                                                                                 \\ \hline
\multicolumn{1}{|c|}{2012}                                                            & \multicolumn{1}{c|}{39.1\%}                                                                                                                                       & \multicolumn{1}{c|}{22.7\%}                                                                                                                           & \multicolumn{1}{c|}{16.0\%}                                                                                                      & \multicolumn{1}{c|}{15.5\%}                                                                                                                      & \multicolumn{1}{c|}{6.7\%}                                                                                                                 \\ \hline
\multicolumn{1}{|c|}{2013}                                                            & \multicolumn{1}{c|}{37.3\%}                                                                                                                                       & \multicolumn{1}{c|}{21.7\%}                                                                                                                           & \multicolumn{1}{c|}{17.3\%}                                                                                                      & \multicolumn{1}{c|}{15.9\%}                                                                                                                      & \multicolumn{1}{c|}{7.8\%}                                                                                                                 \\ \hline
\multicolumn{1}{|c|}{2014}                                                            & \multicolumn{1}{c|}{39.4\%}                                                                                                                                       & \multicolumn{1}{c|}{19.8\%}                                                                                                                           & \multicolumn{1}{c|}{16.0\%}                                                                                                      & \multicolumn{1}{c|}{14.9\%}                                                                                                                      & \multicolumn{1}{c|}{9.9\%}                                                                                                                 \\ \hline
\multicolumn{1}{|c|}{2015}                                                            & \multicolumn{1}{c|}{43.9\%}                                                                                                                                       & \multicolumn{1}{c|}{18.7\%}                                                                                                                           & \multicolumn{1}{c|}{14.2\%}                                                                                                      & \multicolumn{1}{c|}{12.2\%}                                                                                                                      & \multicolumn{1}{c|}{11.0\%}                                                                                                                \\ \hline
\multicolumn{1}{|c|}{2016}                                                            & \multicolumn{1}{c|}{45.3\%}                                                                                                                                       & \multicolumn{1}{c|}{18.5\%}                                                                                                                           & \multicolumn{1}{c|}{13.4\%}                                                                                                      & \multicolumn{1}{c|}{10.4\%}                                                                                                                      & \multicolumn{1}{c|}{12.4\%}                                                                                                                \\ \hline
\multicolumn{1}{|c|}{2017}                                                            & \multicolumn{1}{c|}{47.6\%}                                                                                                                                       & \multicolumn{1}{c|}{17.4\%}                                                                                                                           & \multicolumn{1}{c|}{12.4\%}                                                                                                      & \multicolumn{1}{c|}{10.1\%}                                                                                                                      & \multicolumn{1}{c|}{12.5\%}                                                                                                                \\ \hline
\end{tabular}
}
\end{table}

\subsection{Fine-grained Figure Analysis}\label{result4}

\begin{figure}[hbt!]

    \centering
    \includegraphics[width=.8\linewidth]{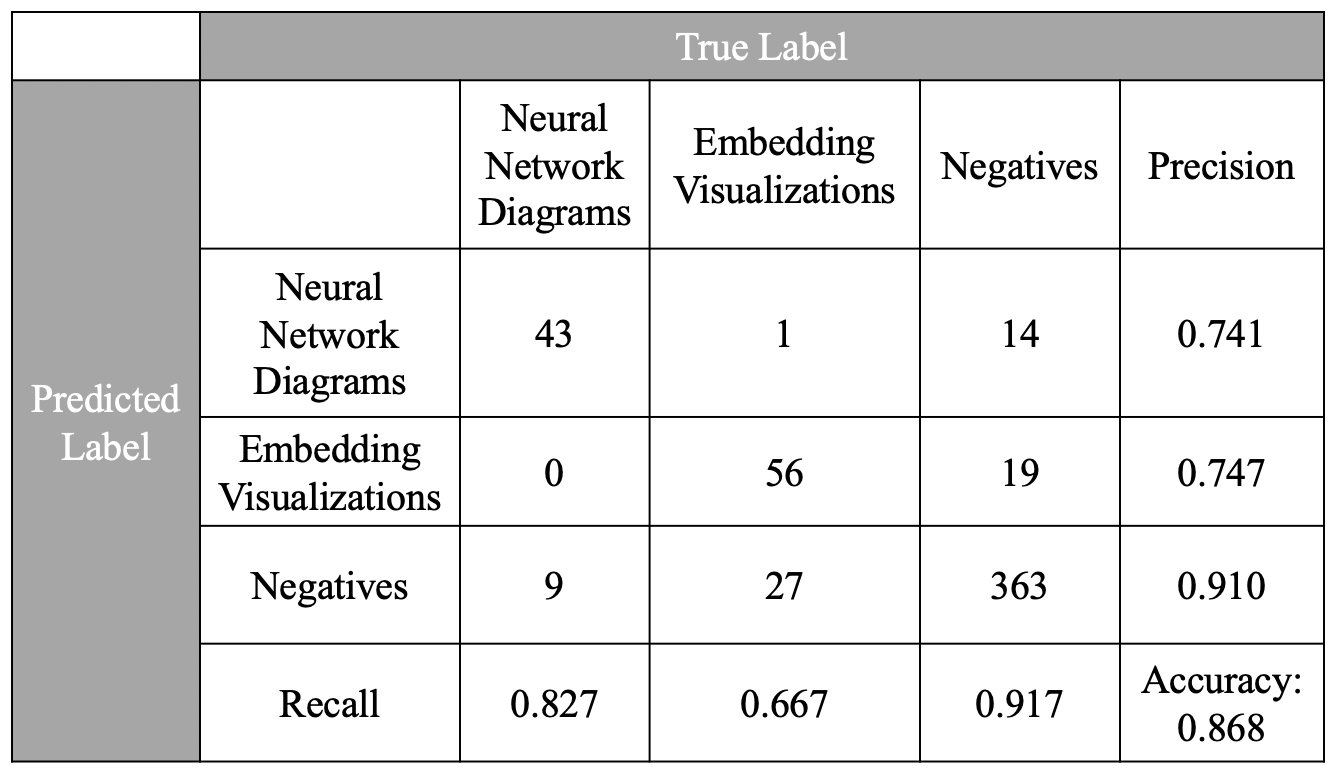}
    \caption{The confusion matrix of the figure type classifier. The classifier achieves 0.868 overall accuracy. }
    \label{table:confusion_matrix}

\end{figure}

The classifier achieves accuracy of 0.902 on the validation set and 0.868 on the test set with precision of 0.741 and recall of 0.827 on neural network diagrams. The confusion matrix of the classifier is shown in Fig. \ref{table:confusion_matrix}. The classifier tends to misclassify flow charts, bar charts, and diagrams with multiple circles as neural network diagrams and the classifier is also often confused between embedding visualization and scatter plots (which are indeed quite similar). The classifier appears sufficiently effective at identifying neural network diagrams and embedding visualizations to conduct following analysis.

\begin{figure}
    \centering
    \includegraphics[ width=.9\columnwidth]{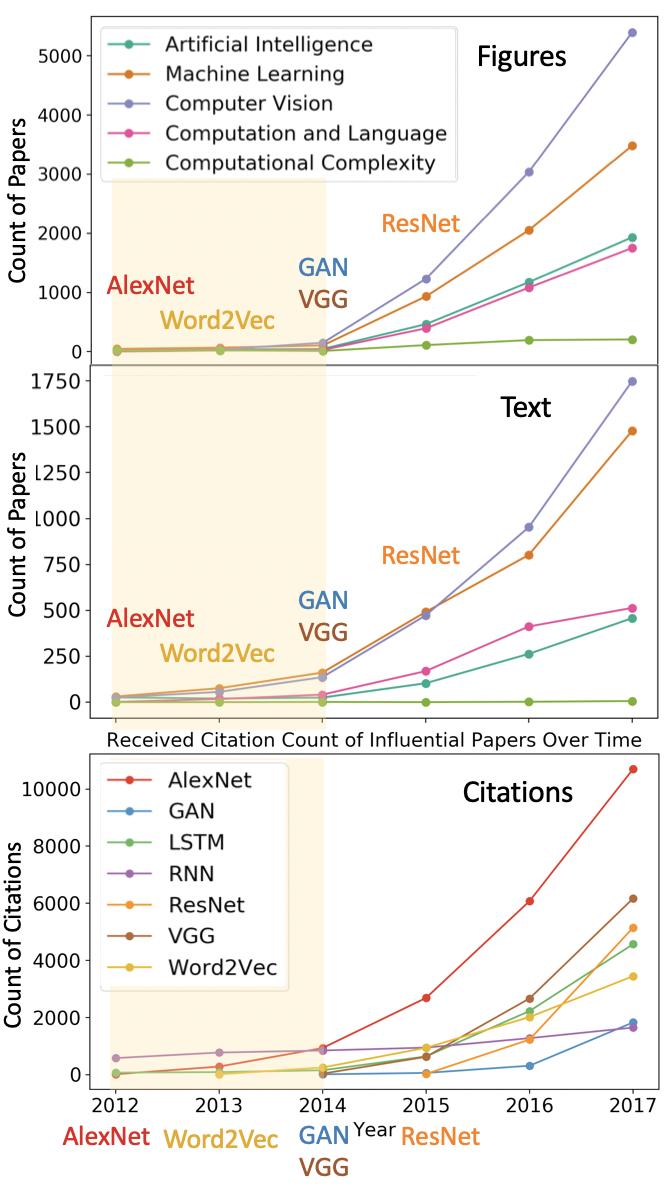}
    \caption{The three line charts demonstrate the trend of recent studies in deep learning using three different media: figures (top), text (middle), and citation (bottom). Top: The number of papers that include neural network diagrams over time. Middle: The count of papers that have "neural network" or "deep learning" in their abstracts over time. Bottom: The citation count of six selected influential papers in deep learning. The annotation of each influential paper indicates the publication time. Citation count of the most influential papers and use of the term "neural network" in the abstract quickly increase (yellow area), but the effect is small.  The use of relevant figures increases only once authors start to truly adopt the concept in their research.}
    \label{fig:stacked_trend}
\end{figure}

We use the trained classifier to label 60k figures in computer science papers on arXiv and analyze the count of the neural network diagrams (Top line chart in Fig. \ref{fig:stacked_trend}) and the embedding visualizations in computer science disciplines over time. We select four categories, which are \textit{Artificial Intelligence}, \textit{Machine Learning}, \textit{Computer Vision}, and \textit{Computation Language}. These disciplines are known to be strongly involved in neural network research. We also include \textit{Computational Complexity}, which has less involvement in neural learning research as a control. We also compute the count of papers whose abstract include "neural network" and "deep learning" in the selected categories over time.  The usage profile by field in the use of embedding visualizations is similar to that of neural network diagrams.  The trend is shown in the middle line chart in Fig. \ref{fig:stacked_trend}. Finally, we select six influential papers in deep learning research: AlexNet \cite{krizhevsky2012imagenet}, GAN \cite{goodfellow2014generative}, LSTM \cite{hochreiter1997long}, ResNet \cite{he2016deep}, RNN \cite{rumelhart1986learning}, VGG \cite{simonyan2014very}, and Word2Vec \cite{mikolov2013distributed}. 
We calculate the received citation count of each paper for each year to show the growth of influence of these papers (Bottom line chart in Fig. \ref{fig:stacked_trend}). We compare these results with our visualization-based metrics to study our third hypothesis: we can use specific types of figures to track the propagation of ideas and methods in the literature.

From the three plots, we make the following observations. First, the three line charts demonstrate the same tendency: a rapid rise in recent years. It is not surprising to see this common trend; increased interest in a topic leads to both increasing citations and an increasing number of relevant diagrams across the literature.

Second, the count of papers that include "neural network" in their abstracts steadily increases from 2012 to 2014 (yellow background), as does the citation count of one particular paper, AlexNet.  But there is no increase in the use of figures during this period.  The cost of mentioning "neural networks" or citing a relevant paper is low, but the cost of developing a relevant figure is high.  We interpret this result as evidence that the use of a figure is better correlated with the true \emph{adoption} of a concept or method, as opposed to simply acknowledging the \emph{relevance} of a concept or method.  After a novel idea is published, the community rapidly begins to discuss the work and, potentially, cites a relevant paper.  But it takes time for the community to integrate the concept into their own research.  Once they have done so, the cost of developing a figure is justified, and the number of figures increases. When the concept is adopting the concept, visualizations begin to emerge in the literature. 

Third, the number of neural network diagrams increases dramatically in 2015 in the four relevant disciplines, while, except for AlexNet, we do not see such rapid growth of received citation counts until 2017 (ResNet and VGG). There is a two year gap between the emergence of the use of neural network diagrams and the rise of the received citation counts.  Figures, as well as text, are faster to react to the introduction of new ideas than aggregate citation counts.  These results both validate the use of figures as a signal of scientific communication, but also that they expose patterns not otherwise discernible.

\section{Conclusion}

In this study, we demonstrate the feasibility of visual information being used as a measure of similarity. We show that visual distance is able to determine the overall relationships between fields by acquiring moderate high correlation (0.706) between visual distance and citation distance. In addition, we show that visual distance still delivers valuable information when it disagrees with citation distance. We further conduct a case study on two specific types of figures: neural network diagrams and embedding visualizations. We find that the upward trend of neural network diagrams and embedding visualizations predates the citation counts of influential papers in recent years. This provides evidence that figures in the scientific literature are leading indicators of citations. 

We plan to extend our study to more fine-grained figure labels. This extension will afford better interpretation of the correlations between figures, text, and citations and help us better refine our groupings. In addition, we plan to apply these visual demarcation techniques to tasks in information retrieval and recommendation systems.

\section{Acknowledgement}
This research has made use of NASA's Astrophysics Data System Bibliographic Services.